\documentclass[conference]{IEEEtran}
\IEEEoverridecommandlockouts
\usepackage{cite}
\usepackage{booktabs}
\usepackage{autobreak}
\usepackage{amsmath,amssymb,amsfonts}
\usepackage{graphicx}
\usepackage{textcomp}
\usepackage{xcolor}

\usepackage{hyperref}
\usepackage{amsmath}
\usepackage{autobreak}
\usepackage{amsfonts}
\usepackage{bm}
\usepackage{bbm}
\usepackage{algorithm}  
\usepackage{algorithmicx}  
\usepackage{algpseudocode}
\usepackage{stfloats}
\usepackage{cuted}
\usepackage{color}
\usepackage{subfigure}
\usepackage{cite}
\usepackage{booktabs} 
\definecolor{red}{RGB}{192,0,0}

\newcommand{\tr}{\text{Tr}}

\newcommand{\hk}{\mathbf{h}_k}

\newcommand{\SINR}{\text{SINR}}
\newcommand{\SCNR}{\text{SCNR}}
\newcommand{\CRB}{Cram\'er-Rao }
\makeatletter
\renewcommand{\maketag@@@}[1]{\hbox{\m@th\normalsize\normalfont#1}}%
\makeatother
\def\BibTeX{{\rm B\kern-.05em{\sc i\kern-.025em b}\kern-.08em
    T\kern-.1667em\lower.7ex\hbox{E}\kern-.125emX}}
\begin{document}

\title{\vspace*{-0.1cm}Flexible Beamforming for Movable Antenna-Enabled Integrated Sensing and Communication}

\author{
	\IEEEauthorblockN{Wanting Lyu$^{*}$, Songjie Yang$^{*}$,Yue Xiu$^{\dag}$, Zhongpei Zhang$^{*}$, \\ Chadi Assi$^{\star}$, \emph{Fellow, IEEE}, and Chau Yuen$^{\ddag}$, \emph{Fellow, IEEE} }
	\IEEEauthorblockA{$^*$ University of Electronic Science and Technology of China, Chengdu, China,}
	\IEEEauthorblockA{$^\dag$ Civil Aviation Flight University of China, Chengdu, China,}
	\IEEEauthorblockA{$^\star$ Concordia University, Montreal, Canada, $^\ddag$ Nanyang Technological University, Singapore.}
	\IEEEauthorblockA{Corresponding author: Zhongpei Zhang (zhangzp@uestc.edu.cn).} 
}

\maketitle
\begin{abstract}
	This paper investigates flexible beamforming design in an integrated sensing and communication (ISAC) network with movable antennas (MAs). A bistatic radar system is integrated into a multi-user multiple-input-single-output (MU-MISO) system, with the base station (BS) equipped with MAs. This enables array response reconfiguration by adjusting the positions of antennas. Thus, a joint beamforming and antenna position optimization problem, namely flexible beamforming, is proposed to maximize communication rate and sensing mutual information (MI). The fractional programming (FP) method is adopted to transform the non-convex objective function, and we alternatively update the beamforming matrix and antenna positions. Karush–Kuhn–Tucker (KKT) conditions are employed to derive the close-form solution of the beamforming matrix, while we propose an efficient search-based projected gradient ascent (SPGA) method to update the antenna positions. Simulation results demonstrate that MAs significantly enhance the ISAC performance when employing our proposed algorithm, achieving a 59.8\% performance gain compared to fixed uniform arrays.

\end{abstract}


\section{Introduction}

Integrated sensing and communication (ISAC) has shown great potential across various applications such as vehicle-to-everything (V2X), industrial internet of things (IIoT), and environment monitoring. By sharing hardware platform and signal processing modules, ISAC enables efficient resource utilization for simultaneous communication and radar sensing \cite{8999605}. The primary goal of ISAC is to enhance both communication capacity and sensing ability. Different metrics have been employed in the literature to evaluate the sensing performance. One is beampattern gain maximization at the target angles, which ensures high beam gain for target sensing \cite{BP1,NOMA2}. Authors in \cite{BP2} proposed a beampattern matching problem, aimed at minimizing the matching error between idealized beampattern and designed one. Considering estimation accuracy, minimizing the \CRB has become another important objective in ISAC beamforming design \cite{CRB1,CCC,CRB2,CRB3}. In more practical cluttered environments, the radar signal-to-clutter-plus-noise-ratio (SCNR) has been exploited \cite{PMN}. Optimizing SCNR is beneficial for enhancing target echoes and suppressing clutter echoes. Furthermore, authors of \cite{SMI2,SMI3} used mutual information (MI) to evaluate the entropy of the received radar signal and maximized MI to in waveform design. 

Recently, movable antennas (MAs), also known as fluid antennas, has been investigated to provide additional degrees of freedom (DoFs) for beamforming \cite{MA-mag,FA}. Compared to conventional MIMO systems, which are limited to optimizing the precoding matrix to enhance channel capacity, MAs offer the flexibility to reconfigure wireless channels. This is achieved by strategically designing the positions of antennas to alter the array response \cite{MA1}. Several works have investigated MAs into communication capacity enhancement \cite{MA2-MIMO,MA-MU4,MA-DOG}. In \cite{MA2-MIMO}, MA multiple-input multiple-output (MIMO) was analyzed, where MAs are utilized at both transmitter and receiver. Results showed that the capacity was significantly increased by antenna positions design.  In \cite{MA-MU4}, MAs are explored in multi-user multiple-input single-output (MU-MISO) system, where gradient ascent (GA) method was used to improve the sum rate of users. In \cite{MA-DOG}, the authors proposed flexible precoding with MAs, which not only adjusted the antenna coefficients (corresponding to traditional precoding) but also optimized element positions under a sparse optimization framework, improving traditional precoding schemes.

Although previous studies have obtained satisfactory results in MA-assisted systems, they focused exclusively on communication. To bridge this gap, this paper explores MAs in ISAC system to simultaneously enhance communication and sensing performance. To the best of the authors' knowledge, this is the first paper to investigate MAs-based ISAC, introducing flexible beamforming that simultaneously optimizes both the beamforming (antenna coefficients) and the antenna positions. For a practical consideration, we study a cluttered environment as a source of sensing interference. Thus, communication rate and sensing mutual information (MI) are derived as performance metrics. Then, we propose a problem to maximize the sum of communication rate and sensing MI by optimizing beamforming and the positions of transmit antennas. 
A fractional programming (FP) based alternating optimization (AO) algorithm is proposed to find the optimal solution. Specifically, we utilize Karush–Kuhn–Tucker (KKT) conditions and direct gradient ascent (DGA) method for solving beamforming and antenna positions, respectively, referred to as DGA-based flexible beamforming. Regarding the significance of initial points in DGA and aiming to fully exploit the feasible moving region, we propose a 3-stage search-based projected gradient ascent (SPGA) method to replace DGA, leading to significant performance improvement.


\section{System Model and Problem Formulation}

\begin{figure}[t]
	\centering	\includegraphics[width=0.7\linewidth]{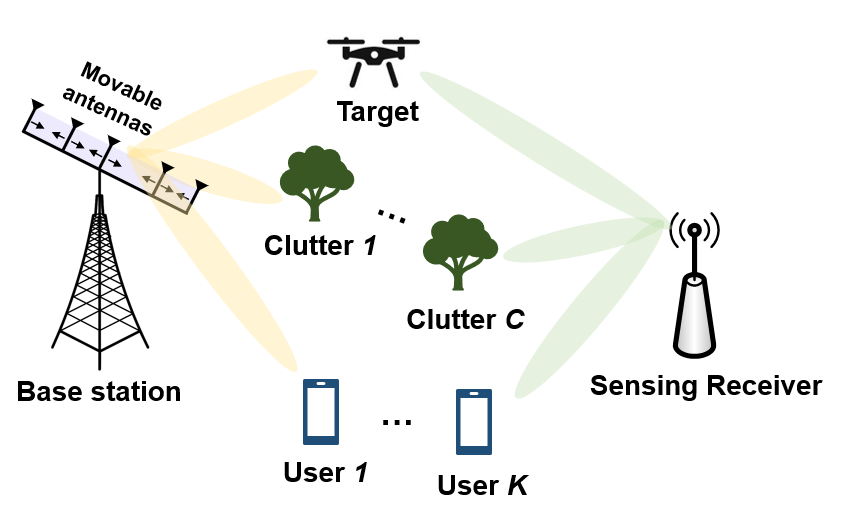}
	\caption{System model of the MA-ISAC system.}
	\label{system model}
\end{figure}

Consider a dual functional radar and communication (DFRC) BS serving $K$ users and sensing one target as shown in Fig. \ref{system model}. Instead of monostatic sensing system, where the transmitter (Tx) and the receiver (Rx) are co-located, we study a bistatic system with separated Tx and Rx. Thus, self-interference can be effectively avoided and ignored \cite{PMN}. In this paper, we consider linear array with MAs. Each user, as well as the sensing Rx are equipped with fixed single antennas.

\subsection{Channel Model}

Based on far-field channel model, the angles of directions (AoDs) of propagation for each antenna element can be regarded as the same. The positions of the transmit MAs are denoted as $\mathbf x = [x_1, \cdots, x_N]^T$. \footnote{Note that when uniform linear array (ULA) with fixed antennas is adopted, $x_n$ can be set as $x_n = (n-1)d_x$, where $d_x$ is the distance of adjacent elements.} Thus, the array response vector of the $l$-th propagation path from the BS to user $k$ is expressed as
\begin{equation}
	\mathbf a_{k,l}(\mathbf x) = \left[e^{j\frac{2\pi}{\lambda}x_1\cos\theta_{k,l}}, \cdots, e^{j\frac{2\pi}{\lambda}x_N\cos\theta_{k,l}} \right]^T \in \mathbb C^{N\times 1}, \label{array_linear}
\end{equation} 
where $\theta_{k,l}$ denotes the angle of direction (AoD) of the $l$-th path of the BS-user $k$ link. Besides, the steering vectors towards the sensing target/clutters are the same as (\ref{array_linear}).

Hence, the communication channel between the Tx and user $k$ can be expressed as
\begin{equation}
	\mathbf h_k(\mathbf x) = \sqrt{\frac{N}{L_p}}\sum_{l=1}^{L_p} \rho_{k,l}\mathbf a_{k,l}(\mathbf x),
\end{equation}
where $\rho_{k,l}$ is the complex channel gain for the $l$-th path of user $k$.

\subsection{Communication and Sensing Signal Model}
We assume the transmit signal to be $\mathbf s = [s_1,\cdots, s_K, s_{K+1}]^T$, where $s_1,\cdots,s_K$ are for communication users $1,\cdots,K$, respectively, and $s_{K+1}$ is dedicated for sensing. Assume that the sensing receiver (SR) has the knowledge of $\mathbf s$, and thus both communication and sensing symbols can be used for target sensing. Without loss of generality, we assume $\mathbf s$ is Gaussian distributed with zero mean and $\mathbb E\left\{\mathbf s\mathbf s^H\right\} = \mathbf I$. The transmit beamforming matrix is denoted as
\begin{equation}
	\mathbf F = [\mathbf f_1, \cdots, \mathbf f_K, \mathbf f_{K+1}] \in \mathbb C^{N\times (K+1)}.
\end{equation}

The channel state information is assumed to be known perfectly by the BS and SR. Based on the channel model, the received signal at user $k$ can be expressed as
\begin{equation}
	y_k = \hk^H(\mathbf x)\mathbf f_k s_k + \sum_{j=1,j\neq k}^{K+1} \hk^H(\mathbf x)\mathbf f_j s_j + n_k,
\end{equation}
where $n_k \sim \mathcal{CN}(0,\sigma_k^2)$ is the additive white Gaussian noise (AWGN). $\mathbf h_k(\mathbf x)$ denotes the channel between the Tx and user $k$ that is related to antenna position $\mathbf x$. The received data rate of user $k$ can be then obtained as
\begin{equation}
	R_k = \log_2(1+\text{SINR}_k),
\end{equation}
where 
\begin{equation}
	\SINR_k = \frac{\left|\hk^H(\mathbf x)\mathbf f_k\right|^2}{\sum_{j=1,j\neq k}^{K+1}\left|\hk^H(\mathbf x)\mathbf f_j\right|^2 +\sigma_k^2}.
\end{equation}

Assume there are $C$ clutters existing as the interference for target sensing. The transmitted symbols are reflected by the sensing target and the clutters, then received by the SR. Thus, the echo signal for sensing can be expressed as
\begin{equation}
	y_s = \alpha_s\mathbf a_s^H(\mathbf x) \mathbf{Fs} + \sum_{c=1}^C \alpha_c\mathbf a_c^H(\mathbf x)\mathbf{Fs} + n_s,
\end{equation}
where $n_s \sim\mathcal{CN}(0,\sigma_s^2)$ denotes the AWGN for radar link. $\alpha_s$ and $\alpha_c$ are complex coefficients including the radar cross section (RCS) of the target/clutter $c$, and cascaded complex gains of the target/clutter $c$, respectively. Additionally, $\mathbf a_s(\mathbf x)$ and $\mathbf a_c(\mathbf x)$ are the array response vectors between the BS and the target/clutter $c$, respectively. 

Thus, we can obtain the radar signal-to-clutter-plus-noise-ratio (SCNR) at the Rx as 
\begin{equation}
	\SCNR = \frac{\left\Vert\alpha_s\mathbf a_s^H(\mathbf x)\mathbf F\right\Vert^2}{\sum_{c=1}^C\left\Vert\alpha_c\mathbf a_c^H(\mathbf x)\mathbf F\right\Vert^2 + \sigma_s^2}.
\end{equation}

According to \cite{SMI2}, the MI can be derived as
\begin{equation}
	R_s = \log_2(1+\SCNR).
\end{equation}

\subsection{Problem Formulation}

In this paper, we aim at maximizing the sum of communication rate and MI. Weighting factors are utilized to control the priority of communication and sensing.

\begin{figure}[t]
	\centering
	\includegraphics[width=0.7\linewidth]{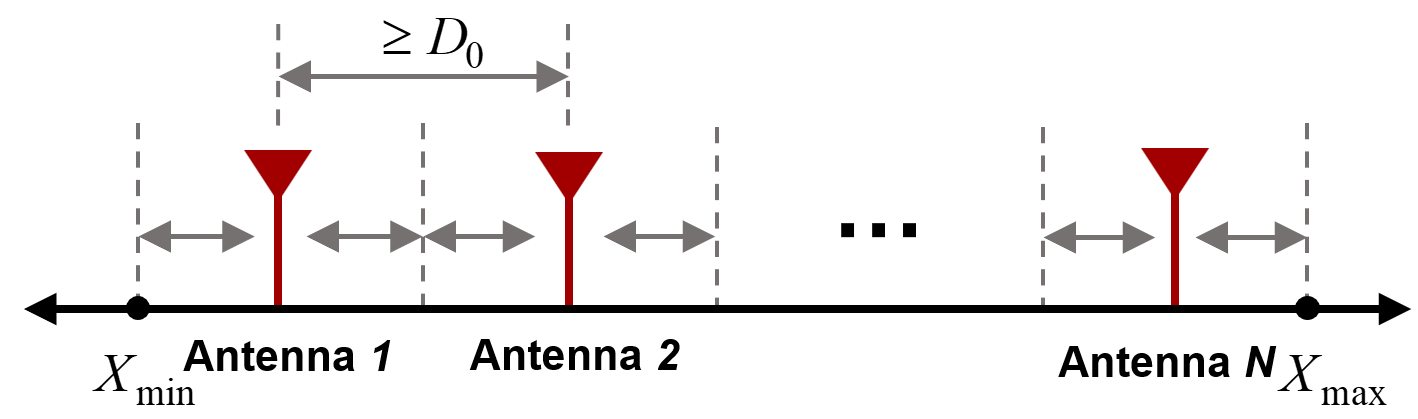}
	\caption{Movable antenna model for linear array.}
	\label{MAmodel}
\end{figure}

As shown in Fig. \ref{MAmodel}, to suppress the coupling effect between the adjacent antenna elements in the flexible array, the antennas need to satisfy a minimum distance constraint:
\begin{equation}
	\left\vert x_z - z_q \right\vert \ge D_0,\; z,q = 1,\dots,N, z\neq q,\label{Cons_dis_linear}
\end{equation}
where $D_0$ denotes the minimum distance between two adjacent elements. Also, the antennas are moving within a feasible region which is predetermined as $\mathcal A = [X_\text{min},X_\text{max}]$.
Accordingly, the flexible beamforming optimization problem can be formulated as
\begin{align}
	(\text{P1})\; \max_{\mathbf F,\mathbf x}\; & \mathcal G(\mathbf F,\mathbf x) = \varpi_c\sum_{k=1}^K R_k + \varpi_sR_s, \label{objfunc} \\
	\text{s. t. } & \tr(\mathbf F^H \mathbf F) \le P_0, \label{Cons_pow} \tag{\ref{objfunc}a}\\
	& \mathbf x \in \mathcal A, \label{Cons_ant_region} \tag{\ref{objfunc}b}  \\
	& (\ref{Cons_dis_linear}), \notag
\end{align}
where $\varpi_c$ and $\varpi_s$ are the weighting factors for communication and sensing, satisfying $\varpi_c + \varpi_s = 1$. (\ref{Cons_pow}) denotes the transmit power budget constraint, $(\ref{Cons_ant_region})$ is the antenna moving region constraint. However, this problem is non-convex with coupled variables. The main difficulty of solving this problem stems from non-convex fractions in objective function, non-convexity of the antenna position constraints, and coupling between beamforming matrix and antenna position vector.

%

\section{Proposed Approach for Linear Arrays}

In this section, we propose an FP-based algorithm to solve the flexible beamforming problem (P1).
We first employ the Lagrangian dual transform and quadratic transform to equivalently transform $\mathcal G(\mathbf F,\mathbf x)$ into (\ref{objFP}) at the bottom of this page, where $\bm\mu = [\mu_1,\cdots,\mu_{K+1}]$, $\bm\xi^c = [\xi_1^c,\cdots,\xi_K^c]^T$ and $\bm\xi^s = [\xi_1^s,\cdots,\xi_{K+1}^s]^T$ are the auxiliary variables. The details of the transformation can be seen from \cite{FP1,PMN}.
\begin{figure*}[b]
	\hrule
	\begin{equation}
		\begin{split}
			&\tilde{\mathcal G}(\mathbf F,\mathbf x,\bm\mu,\bm\xi^c,\bm\xi^s) = \varpi_c\sum_{k=1}^{K}\ln(1+\mu_k)+\varpi_s\ln(1+\mu_{K+1})  - \varpi_c\sum_{k=1}^{K} \mu_k  - \varpi_s\mu_{K+1} \\
			&\qquad+ \varpi_c\sum_{k=1}^K\left(2\sqrt{1+\mu_k}\mathcal Re\left\{ \xi_k^c\hk^H(\mathbf x)\mathbf f_k\right\}  - \vert\xi_k^c\vert^2 \left(\sum_{j=1}^{K+1}\left\vert\hk^H(\mathbf x)\mathbf f_j\right\vert^2 +\sigma_k^2\right)\right) \\
			& + \varpi_s \left( 2\sqrt{1+\mu_{K+1}}\mathcal Re\left\{ \alpha_s\mathbf a_s^H(\mathbf x)\mathbf F\bm\xi^s \right\} - \Vert\bm\xi^s\Vert^2\left( \sum_{c=1}^C\left\Vert\alpha_c\mathbf a_c^H(\mathbf x)\mathbf F\right\Vert^2 + \left\Vert \alpha_s\mathbf a_s^H(\mathbf x)\mathbf F\right\Vert^2 + \sigma_s^2\right) \right).
		\end{split} \label{objFP}
	\end{equation}
\end{figure*}

Since the variables are mutually coupled, we then solve the transformed problem iteratively by AO method. Updating each variable with other fixed variables obtained from the last iteration, we decompose the flexible beamforming problem to four sub-problems as follows.
\begin{itemize}
	\item (SP.1) updating beamforming matrix $\mathbf F$:
	\begin{align}
		\max_\mathbf{F} \;&\tilde{\mathcal G}\left(\mathbf F|\mathbf x^{(t)},\bm\mu^{(t)},\bm\xi^{c(t)},\bm\xi^{s(t)}\right), \\
		\text{s. t. } &(\ref{Cons_pow}). \notag
	\end{align}
	
	\item (SP.2) updating antenna positions $\mathbf x$:
	\begin{align}
		\max_\mathbf{x} \;&\tilde{\mathcal G}\left(\mathbf x|\mathbf F^{(t)},\bm\mu^{(t)},\bm\xi^{c(t)},\bm\xi^{s(t)}\right), \\
		\text{s. t. } &(\ref{Cons_ant_region}), (\ref{Cons_dis_linear}). \notag
	\end{align}
	
	\item (SP.3) updating auxiliary variables $\bm\mu$:
	\begin{equation}
		\max_{\bm\mu} \;\tilde{\mathcal G}\left(\bm\mu|\mathbf F^{(t)}, \mathbf x^{(t)},\bm\xi^{c(t)},\bm\xi^{s(t)}\right).
	\end{equation}
	
	\item (SP.4) updating auxiliary variables $\bm\xi^c$ and $\bm\xi^s$:
	\begin{equation}
		\max_{\bm\xi^c,\bm\xi^s} \;\tilde{\mathcal G}\left(\bm\xi^{c},\bm\xi^{s}|\mathbf F^{(t)}, \mathbf x^{(t)},\bm\mu^{(t)}\right).
	\end{equation}
\end{itemize}

\subsubsection{Updating Transmit Beamforming}

Focusing on (SP.1), to simplify the objective function, we rewrite it as
\begin{equation}
	\begin{split}
		\tilde{\mathcal G}(\mathbf F,\mathbf x,\bm\xi^c,\bm\xi^s) = \sum_{k=1}^{K+1}\big(&2\mathcal Re\{\bm\varphi_k^H\mathbf f_k\} - \mathbf f_k^H\mathbf \Lambda_k\mathbf f_k \big)+ B, \\
		&\forall k\in\{1,...,K+1\},
	\end{split}
\end{equation}
where
\begin{gather}
	\begin{split}
		\bm\varphi_k^{(t)} = \varpi_c\sqrt{1+\mu_k}\xi_k^{c(t)*}\mathbf h_k^T\left(\mathbf x^{(t)}\right)+ \varpi_s\sqrt{1+\mu_{K+1}}\\
		\times\alpha_s^*\xi_k^{s(t)*}\mathbf a_s^T\left(\mathbf x^{(t)}\right),
		\text{ for } K \in \{1,...,K\}, 
	\end{split}\\
	\bm\varphi_{K+1}^{(t)} =  \varpi_s\sqrt{1+\mu_{K+1}}\alpha_s^*\xi_{K+1}^{s(t)*}\mathbf a_s^T\left(\mathbf x^{(t)}\right). \\
	\begin{split}
		\mathbf\Lambda_k = \varpi_c\tilde{\mathbf H}_{k}\tilde{\mathbf H}_{k}^H + \varpi_s\left\Vert\bm\xi^{s(t)}\right\Vert^2\Bigg(\sum_{c=1}^C|\alpha_c|^2\mathbf a_{c}\left(\mathbf x\right)\mathbf a_c^H\left(\mathbf x\right), \\
		+ |\alpha_s|^2\mathbf a_s(\mathbf x)\mathbf a_s^H(\mathbf x)\Bigg), k \in\{1,...,K+1\}, 
	\end{split} \\
	\begin{split}
		&B = \varpi_c\sum_{k=1}^{K}\ln(1+\mu_k)+\varpi_s\ln(1+\mu_{K+1}) \\ 
		&- \varpi_c\sum_{k=1}^{K} \mu_k  - \varpi_s\mu_{K+1}-\varpi_c\sum_{k=1}^K|\xi_k^c|^2\sigma_k^2 - \varpi_s\left\Vert\bm\xi^2\right\Vert^2\sigma_s^2.
	\end{split}
\end{gather}
From the above derivations, $B$ is a constant for (SP.1)), and $\tilde{\mathbf H}_{k} = [\xi^c_1\mathbf h_1, \cdots, \xi^c_K\mathbf h_K]$.

Since $\mathbf\Lambda_k$ is positive definite, (SP.1) is a convex problem that can be efficiently solved by standard solvers such as CVX with high complexity. Then, we propose a low-complexity algorithm solving $\mathbf F$ by deriving the closed-form expression based on Lagrangian dual decomposition method \cite{boyd2004convex}. The Lagrangian of the problem can be first derived as
\begin{equation}
	\mathcal L(\mathbf F, \lambda) = -\tilde{\mathcal G}\left(\mathbf F|\mathbf x^{(t)},\bm\mu^{(t)},\bm\xi^{c(t)},\bm\xi^{s(t)}\right) + \lambda(\tr\left(\mathbf F^H\mathbf F\right) - P_0),
\end{equation}
where $\lambda \ge 0$ is the Lagrangian multiplier corresponding to the power constraint. KKT conditions are used to solve the dual problem as
\begin{small}
\begin{gather}
	\frac{\partial \mathcal L(\mathbf F, \lambda)}{\partial \mathbf F} = \mathbf 0,\label{KKT_stationarity} \\
	\tr\left(\mathbf F^H\mathbf F\right) - P_0 \le 0,\label{KKT_feasibility} \\
	\lambda \ge 0,  \label{KKT_dual_feasibility}\\
	\lambda\left(\tr\left(\mathbf F^H\mathbf F\right) - P_0\right) = 0, \label{KKT_complementary_slackness}
\end{gather}
\end{small}

First, by solving (\ref{KKT_stationarity}), we can obtain the optimal solution of $\mathbf F$ as
\begin{equation}
	\mathbf f_k(\lambda) = \left(\left(\mathbf\Lambda_k^{(t)T} + \lambda\mathbf I\right)^\dagger\right)^*\bm\varphi_k^{(t)}, \;\forall k = \{1,\cdots,K+1\}. \label{update_F}
\end{equation}
The value of $\lambda$ needs to be chosen to satisfy the dual feasibility (\ref{KKT_dual_feasibility}) and the complementary slackness condition (\ref{KKT_complementary_slackness}). If the primal feasibility (\ref{KKT_feasibility}) is satisfied when $\lambda = 0$, the optimal beamforming is $\mathbf f_k(0)$. Otherwise, an appropriate $\lambda$ needs to be decided to satisfy
\begin{equation}
	h(\lambda) = \tr\left(\mathbf F^H(\lambda)\mathbf F(\lambda) \right)- P_0 = 0.
\end{equation}
It can be proved that $h(\lambda)$ is monotonically decreasing with respect to $\lambda$ \cite{JSAC_KKT}. Hence, the bisection method can be adopted to find the solution of $\lambda$, which is summarized in \textbf{Algorithm \ref{bisection}}.

\begin{algorithm}[t]  
	\caption{Bisection Method for Searching dual variable $\lambda$.}
	\begin{algorithmic}[1]  
		\State \textbf{Initialize} upper and lower bound $\lambda_\text{max}$, $\lambda_\text{min}$, tolerance $\varepsilon$ and iteration index $l = 0$.
		\Repeat 
		\State Compute $\lambda^{(l)} = (\lambda_\text{min} + \lambda_\text{max})/2$.
		\State Replace $\lambda^{(l)}$ in $\mathbf F(\lambda)$ and compute $h(\lambda^{(l)})$.
		\State If $h(\lambda^{(l)}) > P_0$, set $\lambda_\text{min} = \lambda^{(l)}$. Otherwise, set $\lambda_\text{max} = \lambda^{(l)}$.
		\State  Set iteration index $l = l+1$.
		\Until{$\vert h(\lambda^{(l)}) - P_0\vert \le \varepsilon$.}
		\State \textbf{Output}: optimal dual variable $\lambda^\star$.
	\end{algorithmic} 
	\label{bisection}
\end{algorithm}

\subsubsection{Updating Antenna Positions}

With fixed beamformer, we update antenna positions $\mathbf x$ by solving (SP.2). Since the objective function $\tilde{\mathcal G}(\mathbf x|\mathbf F^{(t)},\bm\mu^{(t)},\bm\xi^{c(t)},\bm\xi^{s(t)})$ is highly non-convex but differentiable, inspired by \cite{MAMU6}, we propose a 3-stage SPGA algorithm to find a local optimal solution. Since $\tilde{\mathcal G}(\mathbf x|\mathbf F^{(t)},\bm\mu^{(t)},\bm\xi^{c(t)},\bm\xi^{s(t)})$ is highly non-convex, GA is limited to find a local optimum around the initial point. Thus, the PGA algorithm for this problem includes three stages: i) initial point search, ii) gradient ascent updating, and iii) feasibility region projection. Here, we use $\tilde{\mathcal G}$ to replace $ \tilde{\mathcal G}(\mathbf x,\mathbf F,\bm\mu,\bm\xi^c,\bm\xi^s)$ for simplicity.

Specifically, we first set discrete on-grid search point $\mathcal X$ on the feasible region $\mathcal A$. Then, we find initial $x_n$ that maximizes $\tilde{\mathcal G}$ for each antenna:
\begin{equation}
	x_n = \arg\max_{x_n\in \mathcal X}\tilde{\mathcal G},\;n\in\{1,\cdots,N\}.
\end{equation}
Starting at the pre-designed initial point, we perform step ii).

The gradient $\nabla_{\mathbf x} \tilde{\mathcal G}$ can be written as
\begin{equation}
	\nabla_{\mathbf x}\tilde{\mathcal G} = \left[\frac{\partial \tilde{\mathcal G}}{\partial x_1}, \frac{\partial \tilde{\mathcal G}}{\partial x_2}\cdots, \frac{\partial \tilde{\mathcal G}}{\partial x_N} \right]^T.
\end{equation}
The details of computing the derivatives are omitted here.

Hence, with other antennas at fixed positions, $x_n$ can be alternately updated as
\begin{equation}
	x_n^{(i+1)} = x_n^{(i)} + \kappa^t\nabla_{x_n}\tilde{\mathcal G},\label{update_u1}
\end{equation}
where $(i)$ indicates the value obtained from the last iteration in the inner loop for antenna position optimization, $\kappa^t$ denotes the step size for gradient ascent at each iteration. This inner updating lasts until converging to a stationary point.

However, the updated results may not satisfy the position constraints for the MAs. Therefore, the last step is to project the optimized antenna positions into the feasible region \cite{secureMA}. Note that after optimizing $\mathbf x$, the sequential arrangement of the array elements may have perturbations, i.e. not satisfy $x_1 \le x2 \le \cdots \le x_N$. Hence, different from \cite{secureMA}, we rearrange the indices of antenna elements as $\tilde x_m$, where $X_\text{min} \le \tilde x_1 \le \tilde x_2 \le \cdots \le \tilde x_N \le X_\text{max}$, and each $\tilde x_m$ has its corresponding $x_n$. Recalling the constraints (\ref{Cons_ant_region}) and (\ref{Cons_dis_linear}) for linear arrays, we have 
\begin{small}
\begin{equation}
	\begin{cases}
		\tilde{x}_1 \ge X_\text{min}, \\ \tilde x_2 - \tilde x_1 \ge D_0,\\  \ \ \ \ \ \vdots \\\tilde x_N - \tilde x_{N-1} \ge D_0, \\ X_\text{max} \ge \tilde x_N.
	\end{cases}
\end{equation}
\end{small}
Then, it is intuitive to determine the projection function to update $\tilde x_n^\star$ one by one as
\begin{small}
\begin{gather}
	\tilde x_1^{(t+1)} = \left\{\begin{aligned}
		& X_\text{min}, \text{ if } \tilde x_1 < X_\text{min}, \\
		& \tilde x_1, \text{ if } X_\text{min} \le \tilde x_1 \le X_\text{max} - (N-1)D_0, \\
		& X_\text{max} - (N-1)D_0,  \text{ if } \tilde x_1 > X_\text{max} - (N-1)D_0,
	\end{aligned}\right.  \label{project_u1} \\
    \vdots \notag
\end{gather}
\end{small}
\begin{small}
\begin{gather}
	\tilde x_n^{(t+1)} = \left\{\begin{aligned}
		& \tilde x_{n-1} + D_0, \text{ if } \tilde x_{n} < \tilde x_{n-1} + D_0, \\
		& \tilde x_{n}, \text{ if } \tilde x_{n} + D_0 \le \tilde x_{n-1} \le X_\text{max} - (N-n)D_0, \\
		& X_\text{max} - (N-n)D_0,  \text{ if } \tilde x_n > X_\text{max} - (N-n)D_0,
	\end{aligned}\right.  \label{project_u2} \\
     \vdots \notag
\end{gather}
\end{small}
\begin{small}
\begin{equation}
	\tilde x_N^{(t+1)} = \left\{\begin{aligned}
		& \tilde x_{N-1} + D_0, \text{ if } \tilde x_{N} < \tilde x_{N-1} + D_0, \\
		& \tilde x_{N}, \text{ if } \tilde x_{N-1} + D_0 \le \tilde x_{N} \le X_\text{max}, \\
		& X_\text{max},  \text{ if } \tilde x_N > X_\text{max}.
	\end{aligned}\right.  \label{project_u3}
\end{equation}
\end{small}
Finally, simply assign the values of $x_n^{(t+1)}$ by the previous one-one mapping from $\tilde x_n^{(t+1)}$.

\subsubsection{Updating Auxiliary Variable for Lagrangian Dual Transform}

Regarding (SP.3), we update $\bm\mu$ by taking $\frac{\partial \tilde{\mathcal G}\left(\bm\mu|\mathbf F^{(t)}, \mathbf x^{(t)},\bm\xi^{c(t)},\bm\xi^{s(t)}\right)}{\partial \bm\mu} = 0$, which gives
\begin{equation}
	\mu_k^{(t+1)} =
	\frac{R_k^{(t)2} + R_k^{(t)}\sqrt{R_k^{(t)2}+4}}{2}\; k \in \{1,\cdots,K+1\}, \label{update_mu}
\end{equation}
where $R_k^{(t)} = \mathcal Re\left\{\xi_k^{c(t)}\mathbf h_k^{(t)H}\mathbf f_k^{(t)} \right\},\,k = \{1,\cdots,K\}$, $R_{K+1}^{(t)} = \mathcal Re\left\{\alpha_s\mathbf a_s^{(t)H}\mathbf F^{(t)}\bm\xi^{s(t)} \right\}$.

\subsubsection{Updating Auxiliary Variable for Quadratic Transform}

Given fixed $\mathbf F$ and $\mathbf x$, we can update the auxiliary variables $\bm\xi^c$ and $\bm\xi^s$ by solving (SP.4). Because $\tilde{G}(\bm\xi^c,\bm\xi^s|\mathbf F^{(t)},\bm\mu^{(t)},\mathbf x^{(t)})$ is a concave function w.r.t. $\bm\xi^c$ and $\bm\xi^s$ without any constraint, the optimal values can be obtained by solving $\frac{\partial \tilde{G}(\bm\xi^c,\bm\xi^s|\mathbf F^{(t)},\bm\mu^{(t)},\mathbf x^{(t)})}{\partial\bm\xi^c} = 0$ and $\frac{\partial \tilde{G}(\bm\xi^c,\bm\xi^s|\mathbf F^{(t)},\bm\mu^{(t)},\mathbf x^{(t)})}{\partial\bm\xi^s} = 0$. This gives
\begin{gather}
	\xi_k^{c(t+1)} = \frac{\left(\mathbf f_k^{(t)}\right)^H\mathbf h_k\left(\mathbf x^{(t)}\right)}{\sum_{j=1}^{K+1}\left\vert\hk^H\left(\mathbf x^{(t)}\right)\left(\mathbf f_j^{(t)}\right)^H\right\vert^2 +\sigma_k^2}, \label{update_xic}\\
	\bm\xi^{s(t+1)} = \frac{\alpha_s^*\left(\mathbf F^{(t)}\right)^H\mathbf a_s\left(\mathbf x^{(t)}\right)}{ I_R+ \sigma_s^2}, \label{update_xis}
\end{gather}
where $I_R = \sum_{c=1}^C\left\Vert\alpha_c\mathbf a_{c}^H\left(\mathbf x^{(t)}\right)\mathbf F^{(t)}\right\Vert^2 + \left\Vert\alpha_s\mathbf a_s^H\left(\mathbf x^{(t)}\right)\mathbf F^{(t)}\right\Vert^2$.

Based on the above derivations, the overall algorithm for linear  array is summarized in \textbf{Algorithm \ref{OverallAlg}}.
\begin{algorithm}[t]  
	\caption{Proposed Flexible Beamforming Design for Linear Movable Antenna Arrays.}
	\begin{algorithmic}[1]  
		\State \textbf{Initialize} $\mathbf F^{(0)}$, $\mathbf x^{(0)}$, $\bm\xi^{c(0)}$, $\bm\xi^{s(0)}$. Set iteration index $t = 0$.
		\Repeat 
		\State  Update $\mathbf F^{(t+1)}$ by solving (SP.1) as (\ref{update_F}) and \textbf{Algorithm \ref{bisection}}.
		\State  Step i): search for initial starting points.
		\Repeat
		\State  Step ii): alternatively update $x_n^{(i+1)}$ as (\ref{update_u1}).
		\Until {Converges.}
		\State  Step iii): Project the updated $\mathbf x$ into the feasible region to obtain $\mathbf x^{(t+1)}$ as (\ref{project_u1}), (\ref{project_u2}) or (\ref{project_u3}).
		\State Update the auxiliary variable $\bm\mu^{(t+1)}$  as (\ref{update_xic}) and (\ref{update_xis}), respectively.
		\State Update the auxiliary variables $\bm\xi^{c(t+1)}$ and $\bm\xi^{c(t+1)}$ as (\ref{update_mu}).
		\State Update iteration index $t = t+1$.
		\Until{The value of the objective function converges.}
		\State \textbf{Output}: optimal $\mathbf F^\star$, $\mathbf x^\star$.
	\end{algorithmic} 
	\label{OverallAlg}
\end{algorithm}

%

\section{Numerical Results}

In this section, we perform numerical simulations and provide the results to verify the effectiveness and evaluate the performance of our proposed flexible beamforming-ISAC algorithm. 

In the ISAC system, we consider $K=4$ users and $C=3$ clutters. $N=4$ and $N=8$ transmit MAs are set to evaluate the corresponding performance, respectively. The lower bound of the feasible region for MAs is set as $X_{\min} = 0$, while $X_{\max}$ is set as varying parameter to analyze the performance. We set the minimum distance between two adjacent antennas as $D_0 = \frac{\lambda}{2}$. The users and clutters are randomly distributed within the range of $[0,\pi]$, and the target is located at $60^\circ$. $Lp=13$ paths are set for each communication user. The complex channel gain, as well as the complex coefficients for target and clutters follow identical CSCG distribution, i.e. $\rho_{k,l}, \alpha_s,\alpha_c \sim \mathcal{CN}(0,1)$. The noise power for each user and target is assumed to be $\sigma_k^2 = \sigma_s^2 = 1$. The wavelength is $\lambda = 0.1$ m.

Three algorithms are applied in the following simulations, i.e. i) proposed SPGA-based flexible beamforming with MAs (\textbf{SPGA-FBF, MA}), ii) DGA-based flexible beamforming with MAs (\textbf{DGA-FBF, MA}), and iii) beamforming with fixed-position antennas (\textbf{BF, FPA}). For DGA-FBF, direct gradient ascent method is utilized to update antenna positions, until the constraints are not satisfied. The initial setup is as ULA with $\frac{\lambda}{2}$ space between adjacent antennas. For beamforming with fixed antennas, the transmit array is configured as ULA, and optimize beamforming matrix as proposed.

\begin{figure}[t]
	\centering
	\includegraphics[width=0.78\linewidth]{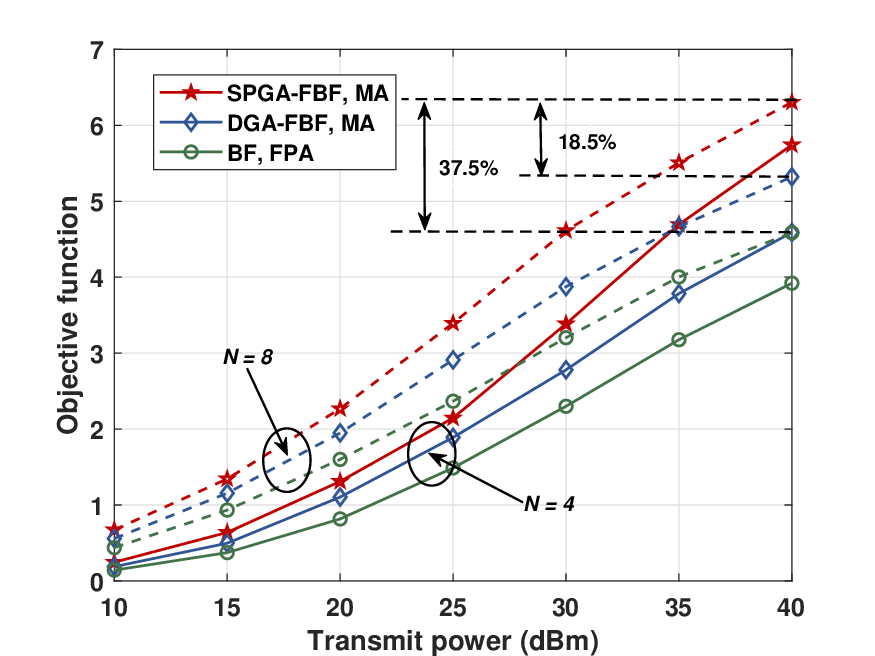}
	\caption{Objective function versus transmit power budget. $X_\text{max} = 10\lambda$, $\varpi_c = 0.5$.}
	\label{OFvsP0}
\end{figure}
\begin{figure}[t]
	\centering
	\includegraphics[width=0.78\linewidth]{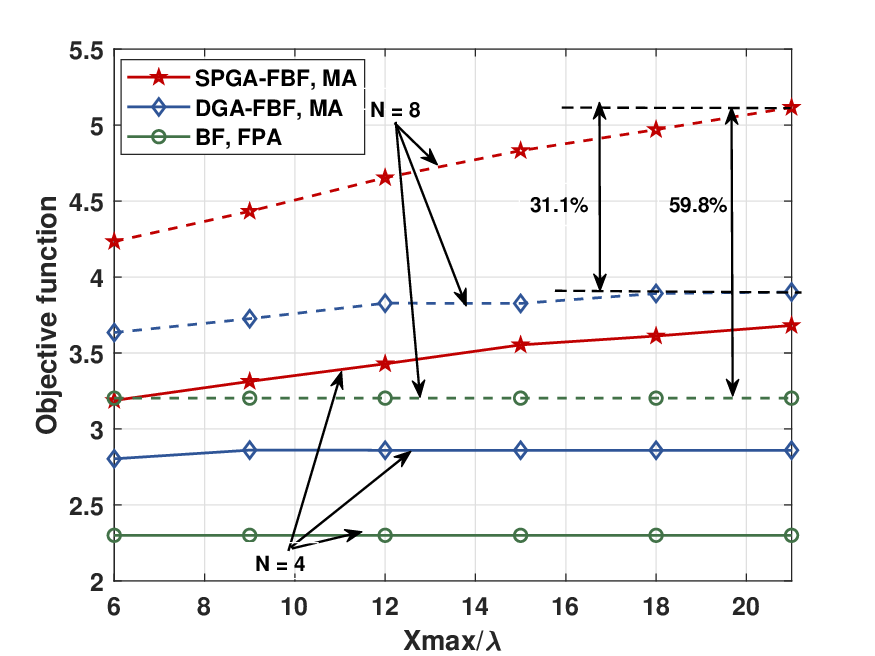}
	\caption{Objective function versus antenna moving region. $P_0 = 30$ dBm, $\varpi_c = 0.5$.}
	\label{OFvsXmax}
\end{figure}
\begin{figure}[t]
	\centering
	\includegraphics[width=0.78\linewidth]{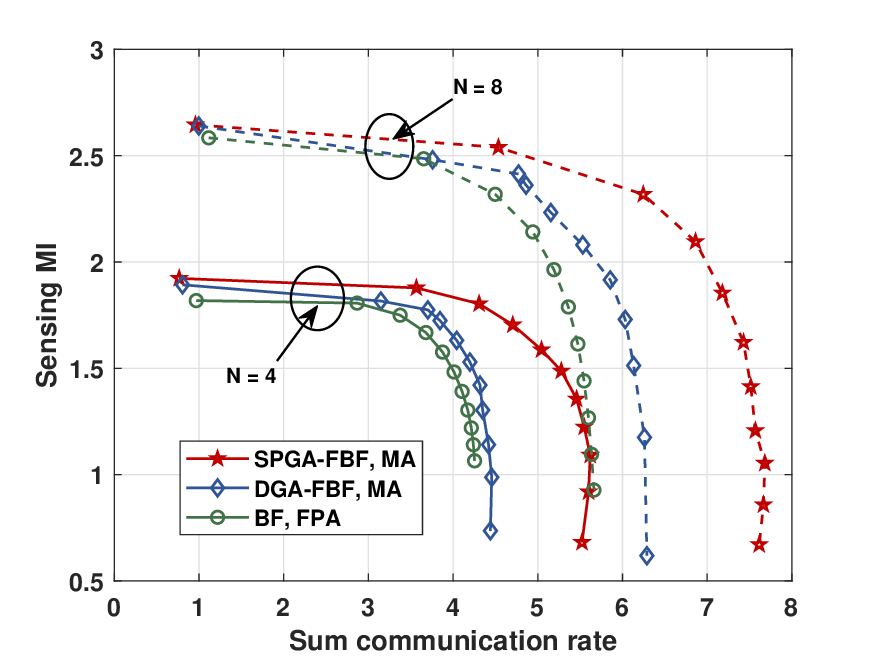}
	\caption{Trade-off between sensing and communication rate. $P_0 = 30$ dBm, $X_{\text{max}} = 10\lambda$.}
	\label{RsvsRc}
\end{figure}

Fig. \ref{OFvsP0} shows how the objective function (sum of rate and MI) varies along increasing transmit power budget $P_0$. The power budget at the BS grows from $10$ to $40$ dBm, which is equivalently to average SNR from $-20$ to $10$ dB. It can be seen that the objective function rises rapidly as transmit power increases, due to stronger received signals at both users and the SR. Furthermore, a larger number of antennas can provide higher beamforming gain, as verified by the superior performance with 8 transmit antennas than $N=4$. From the figure, the proposed MA-based algorithm significantly outperforms the baseline approaches, achieving performance gains of $37.5\%$ and $18.5\%$ compared to fixed antenna configuration and DGA method, in high SNR region.

In Fig. \ref{OFvsXmax}, we analyze the performance variation along with the size of feasible antenna region, by fixing $X_{\text{min}}$ at 0 and varying $X_{\text{max}}$ from $6\lambda$ to $21\lambda$. Significant performance improvement of the proposed algorithm can be observed from the figure, where a larger moving region provides more DoFs to optimize the wireless channel, therefore resulting in a higher value of the objective function. Also, the proposed SPGA-based flexible beamforming significantly outperforms DGA-based flexible beamforming, since the GA method can only find a limited local stationary point without initial search. It is worth noting that with the proposed flexible beamforming algorithm, using only 4 antennas can exceed the performance of the system using 8 fixed antennas, effectively reducing the hardware cost. 

Fig. \ref{RsvsRc} illustrates the trade-off between sensing and communication performance by varying the weighting factor $\varpi_c$ from 0 to 1, which corresponds to $\varpi_s$ changing from 1 to 0. Similarly, we can see that the proposed algorithm performs better than the baselines. The results demonstrate that flexible beamforming with MAs are effective for ISAC systems, and the weighting factor should be carefully selected to meet the specific sensing and communication requirements in practical applications.

\section{Conclusion}

Overall, this paper studied flexible beamforming in an MA-enabled ISAC system. The sum of communication rate and sensing MI was maximized with a weighting factor to control the priority. To tackle the non-convexity of the problem, we transformed the objective function by FP method, and alternatively solved four sub-problems. Finally, numerical results verified the effectiveness of flexible beamforming with MAs in enhancing performance for ISAC system through the proposed algorithm. Particularly, the proposed SPGA-based scheme showed high performance gains with a large feasible moving region and in high SNR settings, compared to the DGA-based method. It is noteworthy that the performance of the ISAC system with only 4 MAs can outperform the system with 8 fixed antennas. This provides insights for reducing the hardware costs in engineering applications.

\ifCLASSOPTIONcaptionsoff
\newpage
\fi

\bibliographystyle{IEEEtran}
\bibliography{Ref}
\end{document}